\documentclass[aps,11pt]{revtex4}
\usepackage{graphicx}

\begin{document}
\title{Superlinear Increase of Photocurrent due to Stimulated Scattering into a Polariton Condensate}
\author{D.M. Myers,$^1$ B. Ozden,$^1$ M. Steger,$^2$ E. Sedov,$^{3,4}$ A. Kavokin,$^{3,5,6}$  K. West,$^7$ L.N. Pfeiffer$^7$ and D.W. Snoke$^{1}$
\\
{\em $^1$Department of Physics and Astronomy, University of Pittsburgh, Pittsburgh, PA 15260, USA\\
$^2$National Renewable Energy Lab, 
Golden, CO 80401, USA\\
$^3$Department of Department of Physics and Astronomy, University of Southampton, Southampton SO17 1BJ, UK\\
$^4$ Department of Physics and Applied Mathematics, Vladimir State University, 87 Gorky str. 600000, Vladimir, Russia\\
$^5$CNR-SPIN, Rome, I-00133, Italy\\
$^6$Spin Optics Laboratory, St.-Petersburg State University, St.-Petersburg 198504, Russia\\
$^7$Department of Electrical Engineering,  Princeton University, Princeton, NJ 08544, USA}}

\begin{abstract}
We show that when a monopolar current is passed through an $n$-$i$-$n$ structure, superlinear photocurrent response occurs when there is a polariton condensate. This is in sharp contrast to the previously observed behavior for a standard semiconductor laser. Theoretical modeling shows that this is due to a stimulated exciton-exciton scattering process in which one exciton relaxes into the condensate, while another one dissociates into an electron-hole pair.  
\end{abstract}

\maketitle

Polariton condensation is now a well established effect \cite{littlewoodreview,carureview,dengreview}.  The polaritons, which can be viewed as dressed photons with effective mass and repulsive interactions, undergo Bose-Einstein condensation and show many of the effects of bosonic stimulation and superfluidity\cite{amo,keeling}, including rapidly flowing to the ground state of whatever potential profile they occupy \cite{pillar}.  Polariton condensates can range from strongly nonequilibrium all the way to equilibrium \cite{prl,sanvitto}. Many previous experiments have focused either on purely optical behavior, in which an optical pump produces optical emission from a polariton condensate; a small number have studied the production of a polariton condensate using electrical injection of free carriers in $p$-$i$-$n$ structures \cite{pin1,pin2,pin3}. In this paper, we show a unique effect of a polariton condensate acting the other way, in which the condensate strongly affects an electrical current. We show that the macroscopic coherence of the polariton wave function dramatically affects the electrical transport, even though the electrical current itself is incoherent. 

\section{Experimental Method}

Figure~1(a) shows the structure of the samples used in our experiments. The central, active region consists of a pillar, typically 100~$\mu$m$~\times~100~\mu$m, which has two distributed-Bragg-reflector (DBR) mirrors, making up an optical cavity. Photons in the microcavity interact with excitons in the quantum wells placed at the antinodes of an optical mode. This part of the structure is the same as that used in previous experiments \cite{pillar,prl,turnaround,gang,mark-prb} which showed long lifetime (~$\gtrsim 200$~ps) and long-distance transport of polaritons (hundreds of microns to millimeters). Our pillar structure is similar to those studied by other groups \cite{pillar2,bloch}, but the long lifetime of the polaritons in our structure allows the polaritons to move across the pillar and find the global potential-energy minima. As reported in Ref.~\cite{pillar} and shown here, in our structures the strain at the edges of the pillar leads to energy minima for the polaritons along the sides which trap the polariton condensate. Figure~1(c) shows a typical intrinsic energy profile for the polaritons, and Figure~1(d) shows the formation of the condensate in these edge traps when the laser light is focused tightly at the center of the pillar, about 40~$\mu$m from where the condensate forms at the edges. As discussed in Ref.~\cite{pillar}, disconnected condensates initially form at the corners of the pillar and then lock to a monoenergetic condensate extending across the pillar as the polariton density is increased. As discussed in Appendix A, we observe three thresholds: one for the onset of quasicondensation at the laser spot, one for ``true''  condensation in localized traps, and one for the phase locking of the separate condensates into a single, global condensate.
\begin{figure}
\includegraphics[width=0.85\textwidth]{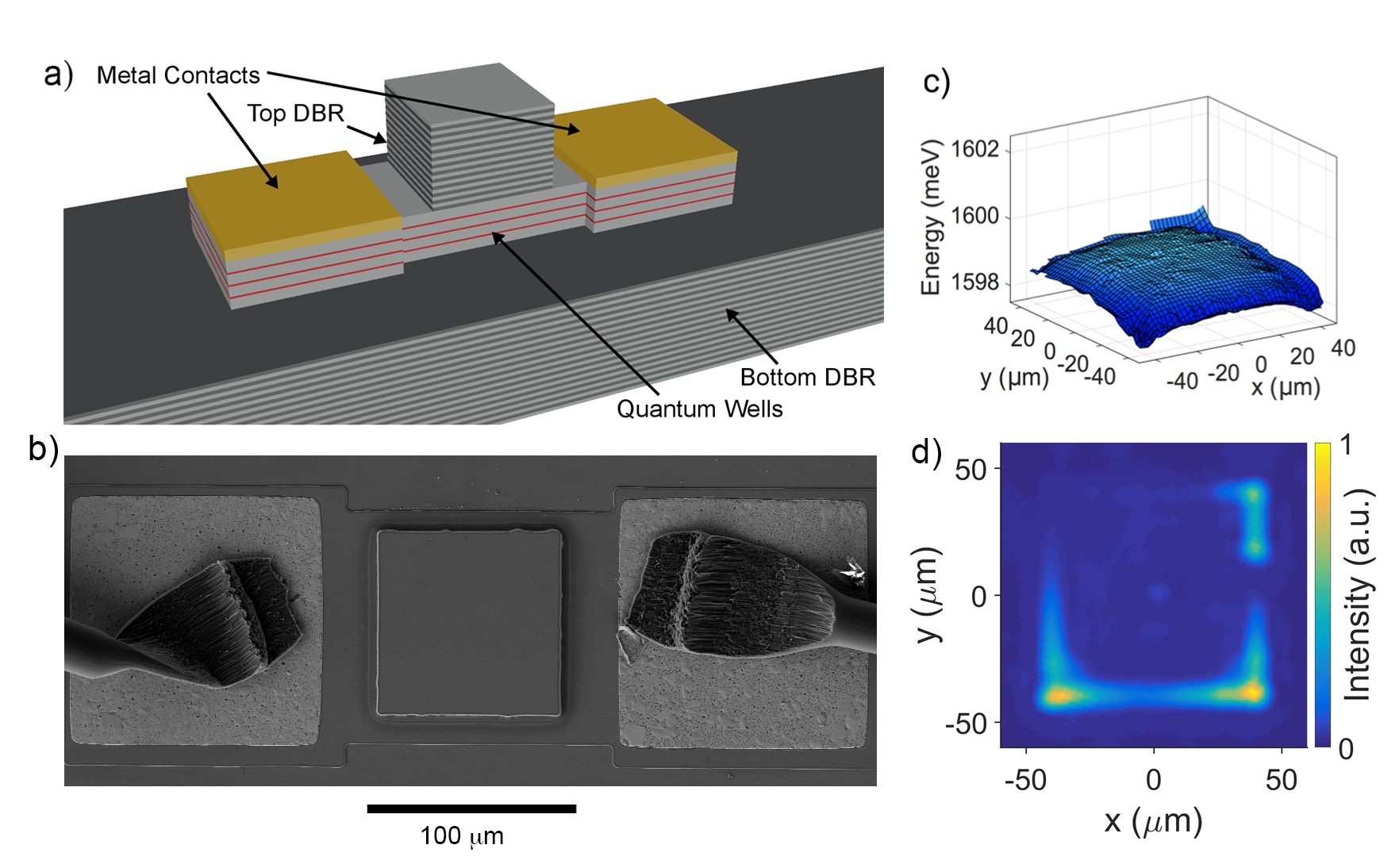}
\caption{a) The structure of the polariton microcavity with symmetric $n$-$i$-$n$ current injection. The distributed Bragg reflector (DBR) mirrors are made of alternating layers of AlAs and Al$_{0.2}$Ga$_{0.8}$As with 40 periods on bottom and 32 periods on top; the QWs are in three sets of four and are pure GaAs/AlAs, and the contacts are both a Ni/AuGe/Ni/Au stack. b) Scanning electron microscope (SEM) image of a typical structure. c) Typical trapping profile of the polaritons in the central pillar structure, obtained by illuminating the sample with a defocused, nonresonant laser at low intensity and using spatially resolved spectroscopy. d) Photoluminescence spatial profile of a condensate in the pillar at high pump intensity (462 mW peak power into a 20~$\mu$m focal spot, chopped with a 2.4\% duty cycle).}
\end{figure}

Outside the pillar, the top mirror is etched away, exposing the quantum wells, and electrical contacts are made to these wells, as illustrated in Figures 1(a) and 1(b).  This allows us to pass free electrons through the polariton condensate in the pillar microcavity. Appendix B gives the measured current-voltage relation of a typical device under various illuminations and a model of the effective circuit.  Previous work has studied the effect of a polariton condensate on photocurrent, e.g. Refs.~\cite{baj1,baj2,winkler}, but in almost all cases with vertical injection; Ref.~\cite{hoefling} also studied lateral injection, similar to our geometry, as discussed below.

When there is no photoexcitation, there is essentially no current through the structure. As the photoexcitation intensity is increased, the current through the structure increases as discussed below. A Keithley 2636B source meter was used to sweep the applied voltage between reverse and forward bias and to measure the current, while the top surface of the microcavity structure was excited by a continuous-wave, stabilized M Squared Ti:Sapphire pump laser with an excess photon energy of about 100 meV, giving an incoherent injection of excitons and polaritons. The pump laser wavelength was varied in each case to give optimal absorption at the lowest dip in reflectivity of the stop band, and the pump laser was chopped to prevent heating of the sample, giving quasi-constant pulses of duration 60~$\mu$s, much longer than the time to reach steady state. Details of the fabrication of the contacts and the synchronization of the electrical and optical measurements are given in Ref.~\onlinecite{photwest}.

\begin{figure}
\protect\includegraphics[width=0.5\textwidth]{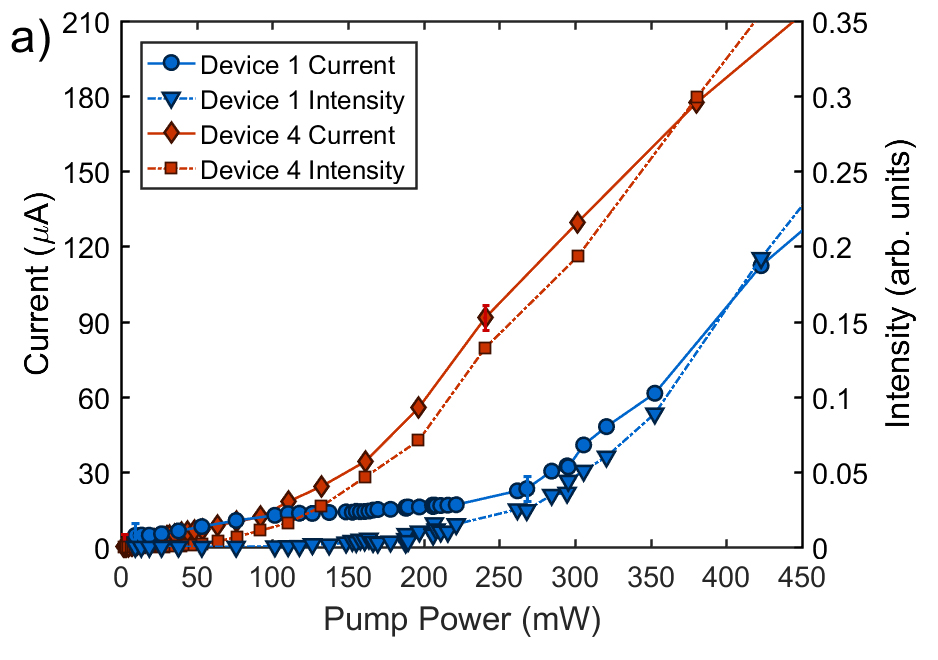}
\protect\includegraphics[width=0.5\textwidth]{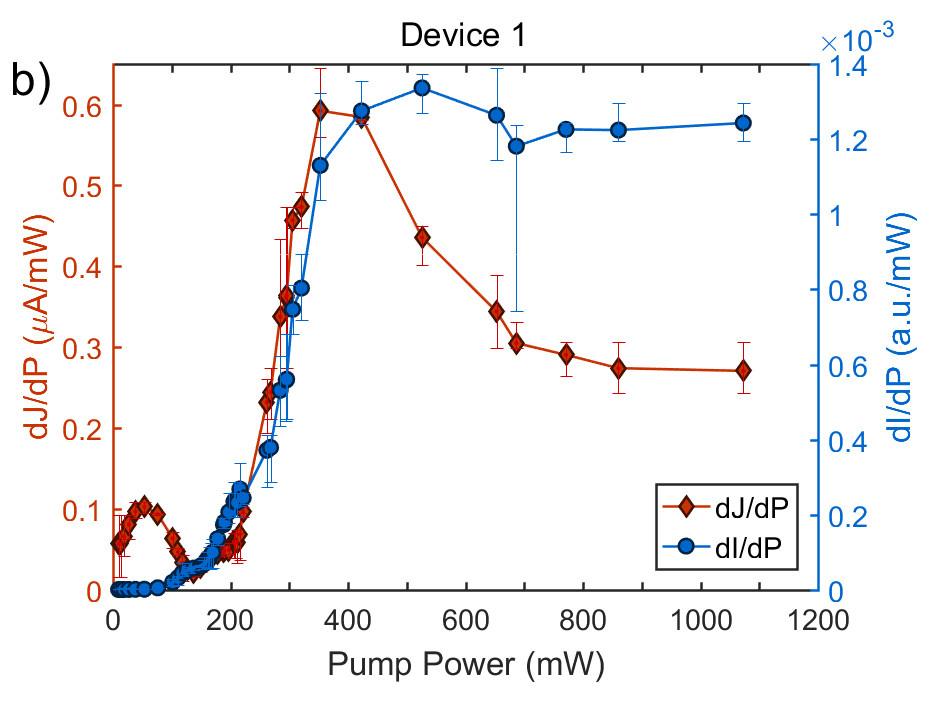}
\caption{a) Solid lines: Current through the structure at constant voltage (1~V) as a function of instantaneous optical pump power, for two different devices, as labeled. The two devices are identical except that they have different condensation threshold densities due to the different photon fractions of the polaritons, controlled by the cavity width. Dashed lines: The emission intensity for the same devices under the same conditions.  The excitation spot size for the two cases was also identical (20 $\mu$m diameter). b) The first derivatives of the data for Device 1 of (a). Red curve, left axis: the first derivative of the current; Blue curve, right axis: the first derivative of the photoluminescence intensity.}
\end{figure}

\section{Experimental Results}

At low optical excitation intensity, the current increases linearly or sublinearly, as is expected if this structure is treated as a standard phototransistor (see Appendix B). When the polariton condensate appears above a critical threshold of optical excitation intensity, however, it has a dramatic effect on the current.  Figure 2(a) shows the current as a function of the optical pump power, at constant voltage, for two different devices.  Linear or sublinear dependence on the illumination power was observed in a very similar structure \cite{hoefling} when there was no extended polariton condensate in the ground state. We see sublinear behavior, as expected from the classical circuit model discussed in Appendix B, when we have a quasicondensate localized at the pump spot, e.g., as seen in the data of Device 1 at low power in Figure 2(a), where a clear saturation is seen. At higher illumination, however, ``true'' condensation occurs, so that we see condensation in local minima far the laser spot, as shown in Figure 1(d) (see also Ref.~\cite{pillar}). (See Appendix A for a discussion of the difference between the ``quasicondensate,''  the ``true'' condensate, and the ``global'' condensation.)  It is only when the ``true'' condensate appears that the current through our structures becomes superlinear with illumination power. This superlinear behavior is seen in every device of the nine we tested; it was always only seen above the critical threshold for ``true'' condensation in the edge traps, and never below this threshold. 

As seen in Figure 2(a), the onset of the superlinear current increase correlates well to the onset of condensation as seen in the light emission from each device.  The critical threshold for condensation was different in different devices because the photon fraction was different in the different devices. This photonic fraction depends on the detuning of the photon cavity mode relative to the quantum well exciton energy, which in turn is determined by the mirror separation in the optical cavity, which varies across the wafer on which the devices were fabricated; the detuning was $\delta = -10.1$~meV for Device 1 and $\delta = -8.2$~meV for Device 4. As seen in this figure, at low power, below 150 mW,  the injected current first increases linearly with the illumination and then saturates, as expected for a classical photodiode. Then when the condensate appears in the corner traps of the pillar, the current jumps up again. Figure 2(b) gives the first derivative of the same data as shown in Figure 2(a), for Device 1, showing that the onset of the jump in current is exactly correlated with the jump in the polariton intensity, i.e., the appearance of the polariton condensate.

\begin{figure*}
\includegraphics[width=0.5\linewidth]{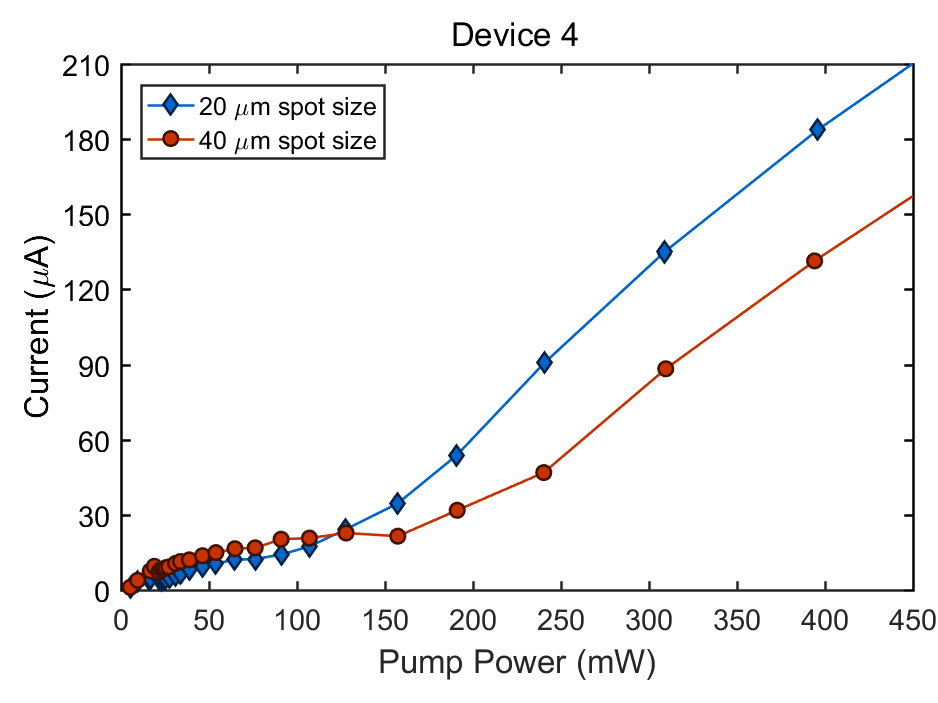}
\caption{ \label{spotsize} Experimental current as a function of pump power for Device 4, for two different pump spot sizes (full width at half maximum), for an appplied voltage of 1~V. 
}
\end{figure*}

\begin{figure*}
\includegraphics[width=0.45\linewidth]{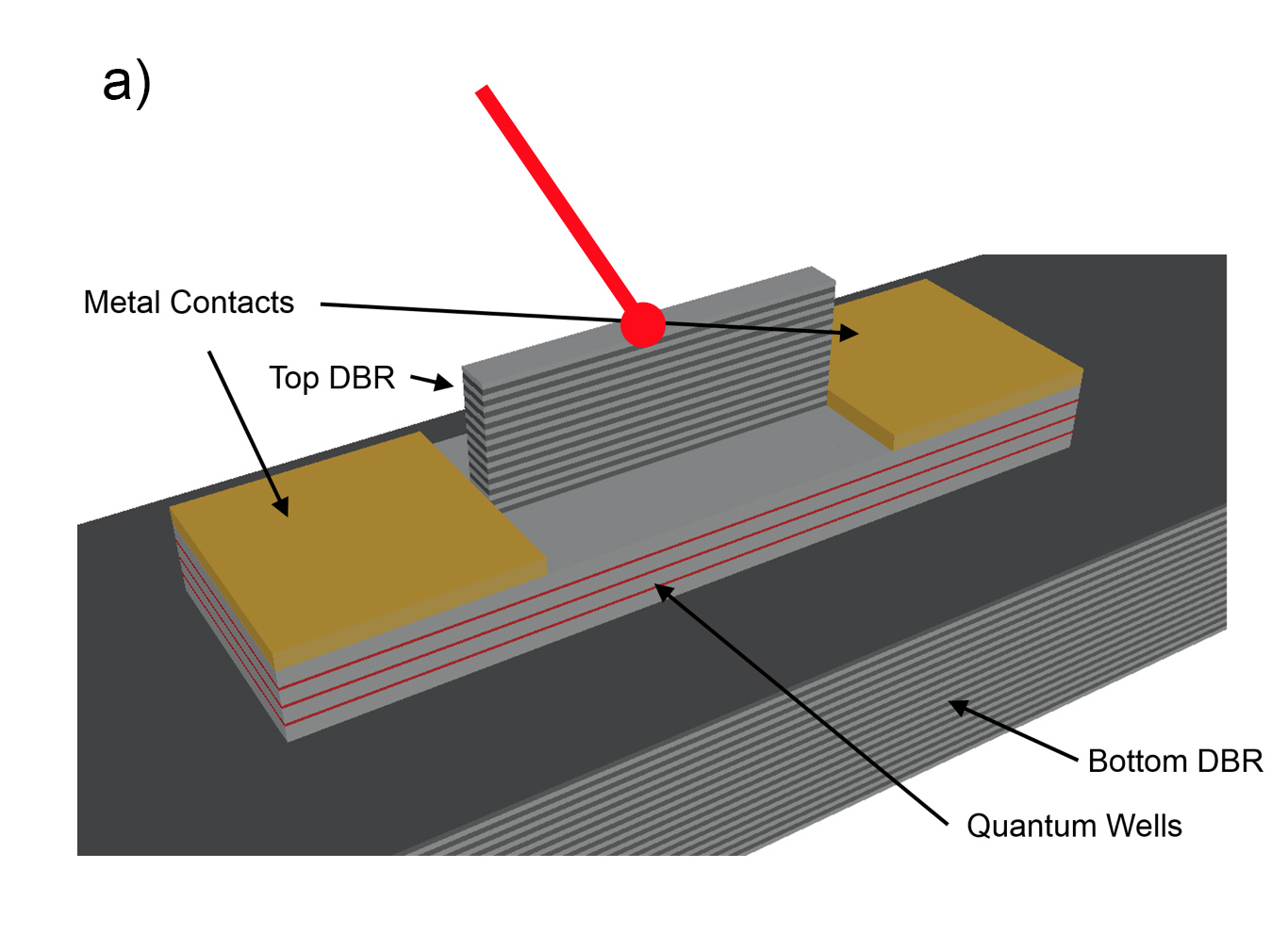}

\includegraphics[width=0.5\linewidth]{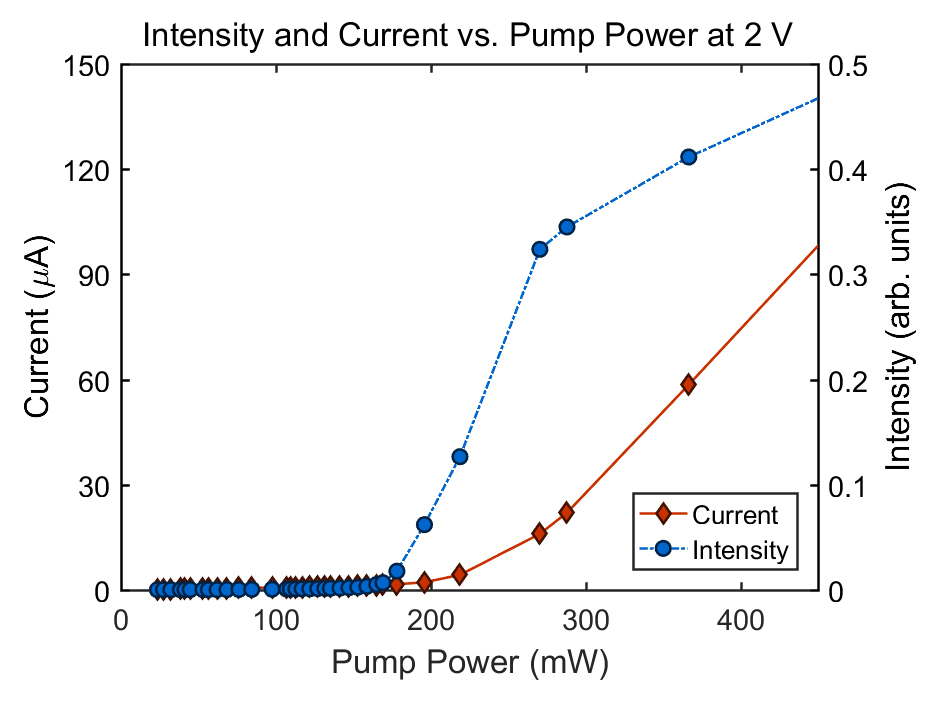}
\includegraphics[width=0.5\linewidth]{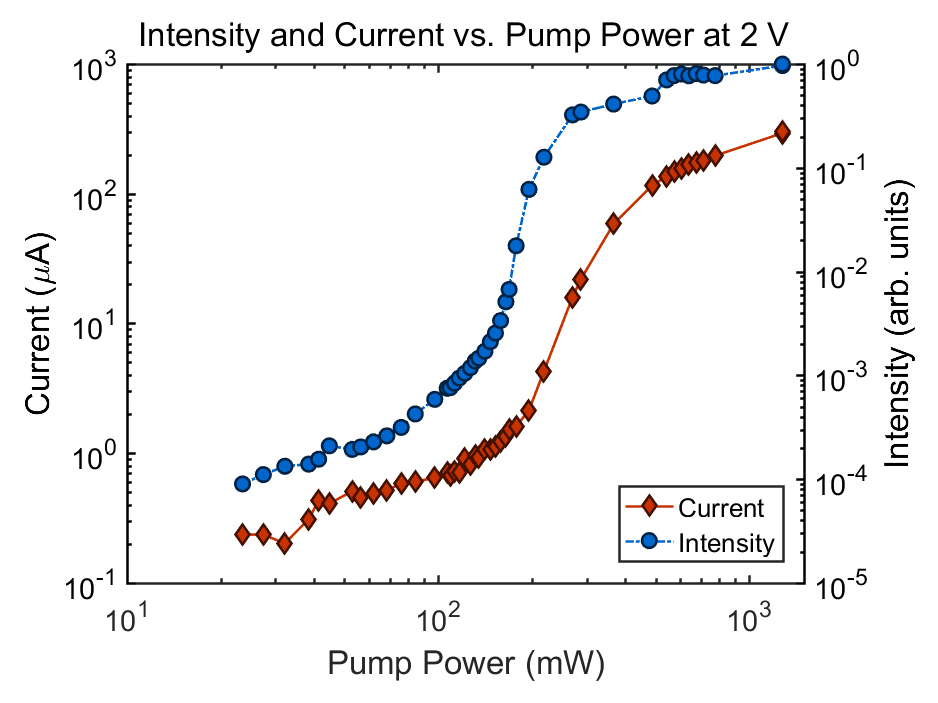}
\caption{ \label{wire} a) Illustration of the wire geometry, fabricated by the same method as the pillars presented in Figure~1. The total length of the wire is 190 $\mu$m, while the width is appriximately 10~$\mu$m.  b)  Linear, and c) logarithmic plot of the current as a function of pump power for this device, for a total voltage drop of 2~V, for a pump spot of approximately 20~$\mu$m in the center of the wire. The detuning of the device was approximately $\delta = -5$~meV.}
\end{figure*}

Figures \ref{spotsize} and \ref{wire} explore the dependence of the effect on the geometry. In Figure~\ref{spotsize}, we compare the current vs. pump power characteristics for two different spot sizes. When the pump spot is smaller, a higher density of the polaritons is created at the center for a given pump power, leading to a lower threshold for the total pump power. This behavior is well reproduced by the theoretical model presented below. 

In Figure~\ref{wire}, we present data from a long wire, created in the same way as the pillars discussed above, but with much larger aspect ratio. As seen in this Figure, the same general behavior applies, with a superlinear increase of the current above the critical threshold for polariton condensation, but in this case there is an offset of the jump of the current to higher pump power, by about a factor of 2. We interpret this as due to the need for the condensate, which is trapped at the ends of the wire at low density, to grow as power increases until it substantially overlaps the edges of the exciton reservoir, which is at the pump spot. As discussed below, our theoretical model indicates that the superlinear increase of the current occurs due to a stimulated process in which some excitons are ionized, creating free electron-hole pairs that contribute to the photocurrent, when the condensate spatially overlaps the exciton reservoir.

\section{Theoretical Model}

To model this result, we have used the following set of equations, which describe the behavior of the coupled three-component system of polaritons, a free exciton ``reservoir,'' and free carriers, similar to the formalism used in Ref.~\onlinecite{bhattkav}:
\begin{eqnarray}
\label{EqGPE}
i \hbar \frac{\partial \Psi}{\partial t} &=&
\left[
-\frac{\hbar ^2}{2 m^{*}} \nabla ^2+ V_{\text{eff}}(t,\mathbf{r}) \right] \Psi + \frac{i \hbar}{2} \left(  R n_{\text{R}}   + { A n_{\text{R}} ^2}   -   { \gamma _{\text{C}} } \right)   \Psi, \\
\label{EqHotExc}
\frac{\partial n_{\text{R}}}{\partial t} &=& { P(\mathbf{r})} -  \left(   R n_{\text{C}}   +   2 A  n_{\text{C}} n_{\text{R}}   
+  \gamma _{\text{R}}  \right)    n_{\text{R}},\\
\label{EqFreCarr}
\frac{\partial n_{eh}}{\partial t} &=&  { A n_{\text{R}} ^2 n_{C} }   - { \gamma _{eh}  n_{eh}} .
\end{eqnarray}
Here Eq.~(\ref{EqGPE}) is the generalized Gross-Pitaevskii equation for the multi-particle wave function of polariton condensate $\Psi (t, \mathbf{r})$,
Equation~(\ref{EqHotExc}) is the rate equation for the density of the reservoir of excitons $n_{\text{R}} (t, \mathbf{r})$, which have higher energy than the polaritons, and do not emit light; Equation~(\ref{EqFreCarr}) describes a reservoir of free carriers (electrons and holes) of density $n_{eh} (t,\mathbf{r})$. In Equation~(\ref{EqGPE}), $m^{*}$ is the effective mass of polaritons, and $V_{\text{eff}}(t,\mathbf{r}) = V(\mathbf{r}) +  \alpha _{\text{C}} n_{\text{C}}  + \alpha _{\text{R}} n_{\text{R}}$ is the effective potential, where $V(\mathbf{r}) $ is the stationary confinement potential across the sample including the effects of strain, cavity gradient, etc. The remaining terms in $V_{\text{eff}}(t,\mathbf{r})$ appear as a result of the polariton-polariton interactions and interaction of the polaritons with the reservoir of hot excitons. The parameter $\alpha _{\text{C}}$ is the polariton interaction constant, $\alpha _{\text{R}}$ is the condensate-reservoir interaction constant, and $n_{\text{C}} = |\Psi|^2$ is the polariton BEC density. The decay lifetimes of the various populations are accounted for by the rates $\gamma_C$, $\gamma_R$, and $\gamma_{eh}$, for the polaritons, excitons, and electron-hole population, respectively. In the case of the free carriers, this can include escape of the carriers through the sides of the device.

A Gross-Pitaevskii equation for the light-mass polariton condensate coupled to a rate equation for the heavy, incoherent exciton reservoir has been used many times to model polariton dynamics, e.g. Refs.~\onlinecite{res1}---\onlinecite{res3}. 
What is new here is accounting explicitly for a free electron-hole population, which is generated by an Auger-like process in which two excitons collide, one scatters down into a polariton state, and the other is ionized to become a free electron-hole pair. This process is described by the terms with the rate constant $A$ in Equations~(\ref{EqGPE})---(\ref{EqFreCarr}). We assume that the optical generation term $P(r)$ generates primarily excitons, due to quick energy relaxation of the photoexcited carriers, and that the primary means of generating free carriers in steady state is collisional ionization of excitons. This process conserves energy, as the energy released during an exciton relaxation from the reservoir to the condensate is used to ionize another exciton. Moreover, this process is stimulated by the occupation number of the polariton condensate, which is why its efficiency strongly increases above the condensation threshold. We also include a term with relaxation rate $R$ for stimulated cooling of excitons into the condensate, e.g., via phonon emission.

For the steady state, from~(\ref{EqGPE})--(\ref{EqFreCarr}) we obtain
\begin{eqnarray}
\label{EqGPESt}
\mu \Psi &=&
\left[
-\frac{\hbar ^2}{2 m^{*}} \nabla ^2+ V_{\text{eff}}(\mathbf{r}) \right] \Psi + \frac{i \hbar}{2} \left( { R n_{\text{R}} } + { A n_{\text{R}} ^2}   -   { \gamma _{\text{C}} } \right)   \Psi,
\\
\label{EqHotExcSt} 
{  P(\mathbf{r})} &=&  \left(  {  R n_{\text{C}}  }
+  { A n_{\text{R}}   }
+ { \gamma _{\text{R}}  }
\right)    n_{\text{R}},
\\
\label{EqFreCarrSt}
n_{eh} &=&  \frac{ {  A n_{\text{R}} ^2 n_{C} }  }{ { \gamma _{eh} } }.
\end{eqnarray}

For the numerical calculations, we used the following values of the parameters: 
$ m ^{*} = 5 \cdot 10 ^{-5} m_e$, where $m_e$ is the free electron mass, ${\gamma _{\text{C}} = 0.004 \, \text{ps}^{-1}}$,
${\gamma _{\text{R}} = 0.04 \, \text{ps}^{-1} }$,
${\gamma _{eh} = 0.6 \, \text{ns}^{-1}}$,
${R = 0.025 \, \text{ps}^{-1}} \cdot \mu\text{m}^2$,
${A = 10^{-15} \, \text{ps}^{-1} \cdot \mu \text{m}^4}$,
${\alpha _{\text{C}} = 0.6 \, \mu\text{eV} \cdot \mu\text{m}^2}$,
${\alpha _{\text{R}} = 2 \alpha _{\text{C}}}$.
The pump is taken to have the Gaussian form $P(\mathbf{r}) \propto \exp \left[ -r^2 / w_{\text{p}}^2 \right]$ with the width $w_{\text{p}} = 10 \, \mu \text{m}$.

\begin{figure*}
\includegraphics[width=0.8\linewidth]{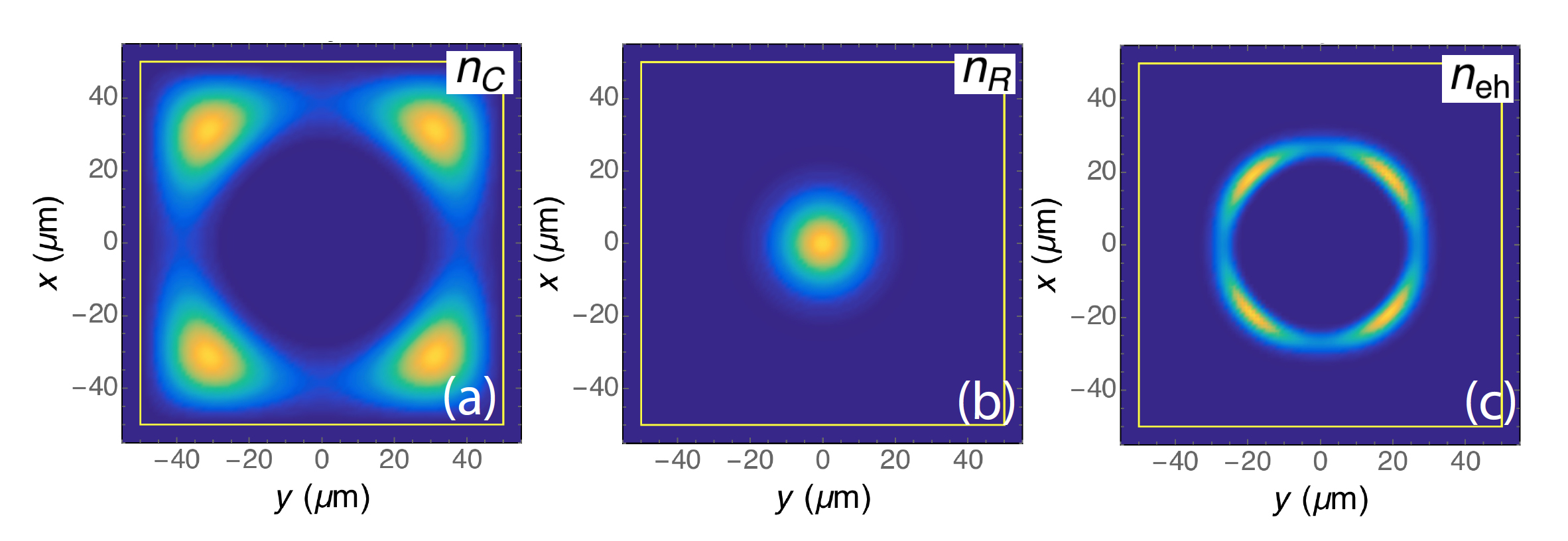}
\caption{ \label{Fig1}
An example of the densities of the polariton BEC (a),
of the reservoir of hot excitons (b) and of the reservoir of free carriers~(c) in the steady state regime.
The yellow rectangle marks boundaries of the trap.
}
\end{figure*}

Figure~3 shows the densities of the polariton BEC, of the exciton reservoir and of the free carrier reservoir in the steady state regime corresponding to the solution of Eqs.~\ref{EqGPESt}--\ref{EqFreCarrSt}, for a square pillar like that shown in Figure 1.
Polaritons collect in the corners of the rectangular trap, as seen in the data of Figure~1(c), because of the polariton flow from the center of the trap where the optical pump is located. Repulsive polariton-polariton and polariton-exciton interactions contribute to the potential energy profile felt by the polaritons as well. 

The exciton reservoir remains localized under the optical pump beam; we ignore exciton diffusion.
According to Eq.~\ref{EqFreCarrSt}, the cloud of free carriers appears at the region of overlap of the condensate and the exciton reservoir, $n_{eh} \propto n_{\text{R}} ^2 n_{\text{C}}$, as seen in the result of the model in Fig.~3(c). 

Figure~4(a) demonstrates the spatially integrated population of the polariton condensate $I_{\text{C}} = \int n_{\text{C}} d\mathbf{r}$ and the total population of the free carrier reservoir $I_{eh} = \int n_{eh} d\mathbf{r}$ in the steady state regime as functions of the optical pump power.
These can be compared to the data shown in Figure 2(a), where the total emission intensity from the condensate is proportional to $I_C$, and the total photocurrent is proportional to the number of free carriers $I_{eh}$. As seen in this figure, the main results of the experiments are reproduced, namely, the polariton condensate population and the photocurrent track with each other, and both show a superlinear increase due to the stimulated collisional term proportional to $A$ in Equations~(\ref{EqGPE})--(\ref{EqFreCarr}). 

\begin{figure*}
\includegraphics[width=0.65\linewidth]{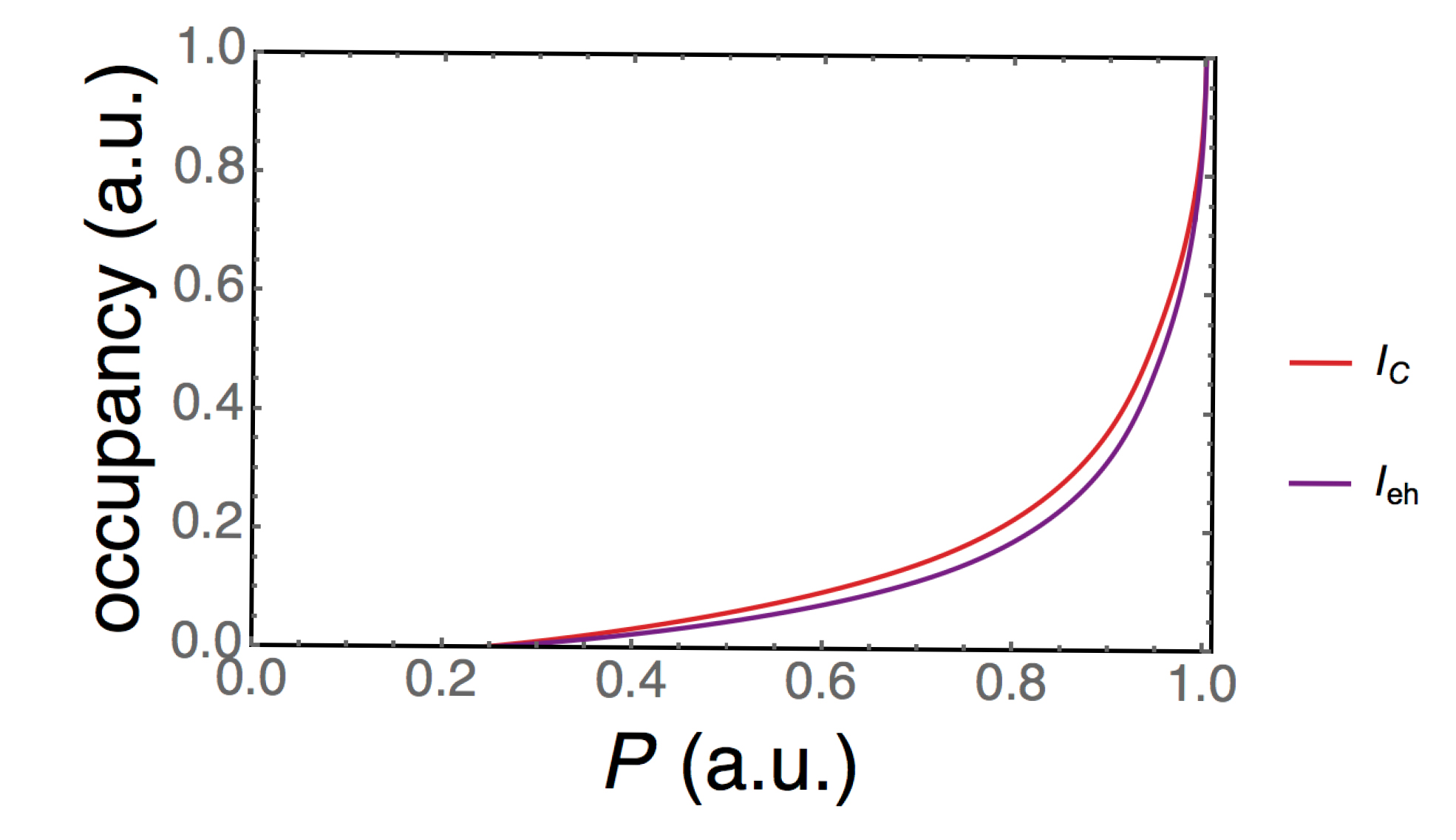}
\caption{ \label{Fig2}
Occupancy of the polariton condensate, $I_{\text{C}} / \max{ (I_{\text{C}})}$, (solid) and the reservoir of the free carriers,  $I_{eh} / \max{ (I_{eh})}$, (dashed) as functions of the optical pump power in the steady state regime.
}
\end{figure*}

\section{Conclusions}

We have seen that the system of a polariton condensate coupled to in-plane current is a highly nonlinear electro-optical device, in which the polariton condensate has a direct effect on the transport due to the existence of stimulated scattering terms that enhance the rate of exciton ionization and therefore increase the population of photogenerated free carriers. The coherence of the condensate leads to the superlinear increase of the current through the structure.  Secondarily, the extremely light polariton mass and the coherent wave function of the polariton condensate leads to an effective transport of polaritons throughout the structure, modeled by the Gross-Pitaevskii equation, which then in turn leads to stimulated creation of free electrons and holes well away from the optical excitation region. 

This new physical effect may potentially have application as a new type of phototransistor with superlinear response. Polariton condensation is now moving toward room temperature \cite{PT2017} and lower threshold \cite{steger-thres}, so that low-power, room-temperature devices using this effect should be possible \cite{bhatt}. Moreover, the experimental set-up described here may be promissing for the observation of the traces of superconductivity mediated by a Bose-Einstein condensate of exciton-polaritons \cite{kavsc}.

{\bf Acknowledgements}. This work has been supported by the Army Research Office Project W911NF-15-1-0466. We thank Allan Macdonald and Ming Xie for conversations on the interaction of electrical transport and optical condensates, and Yun-Yi Pai and Dengyu Yang for assistance on Hall measurements of the carrier density in the devices.

\appendix

\section{Three-threshold behavior}
\begin{figure}
\begin{center}
\includegraphics[width=0.5\textwidth]{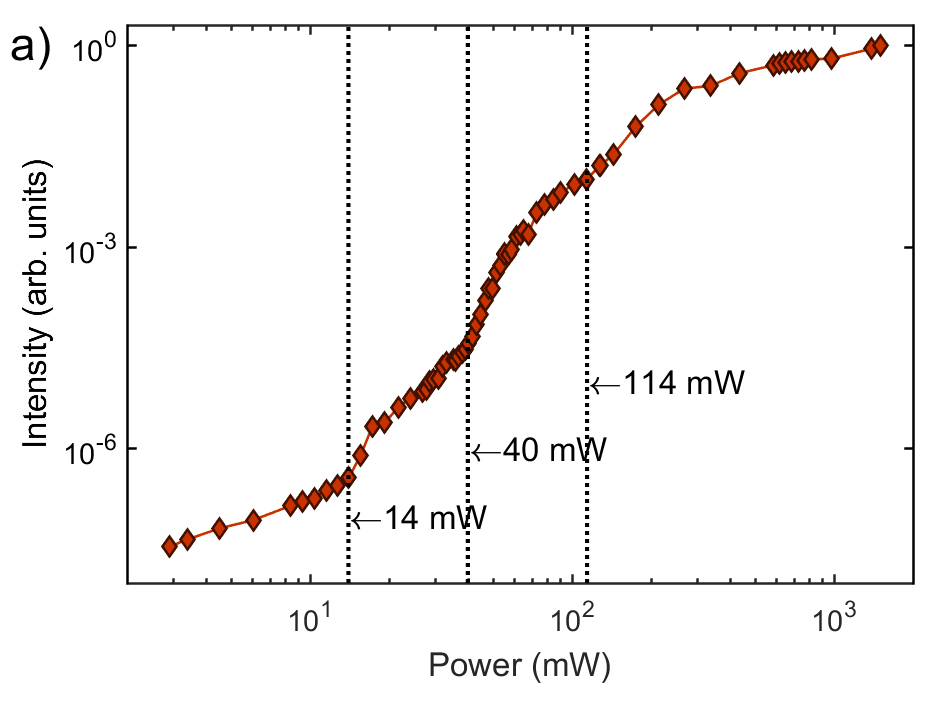}
\includegraphics[width=0.5\textwidth]{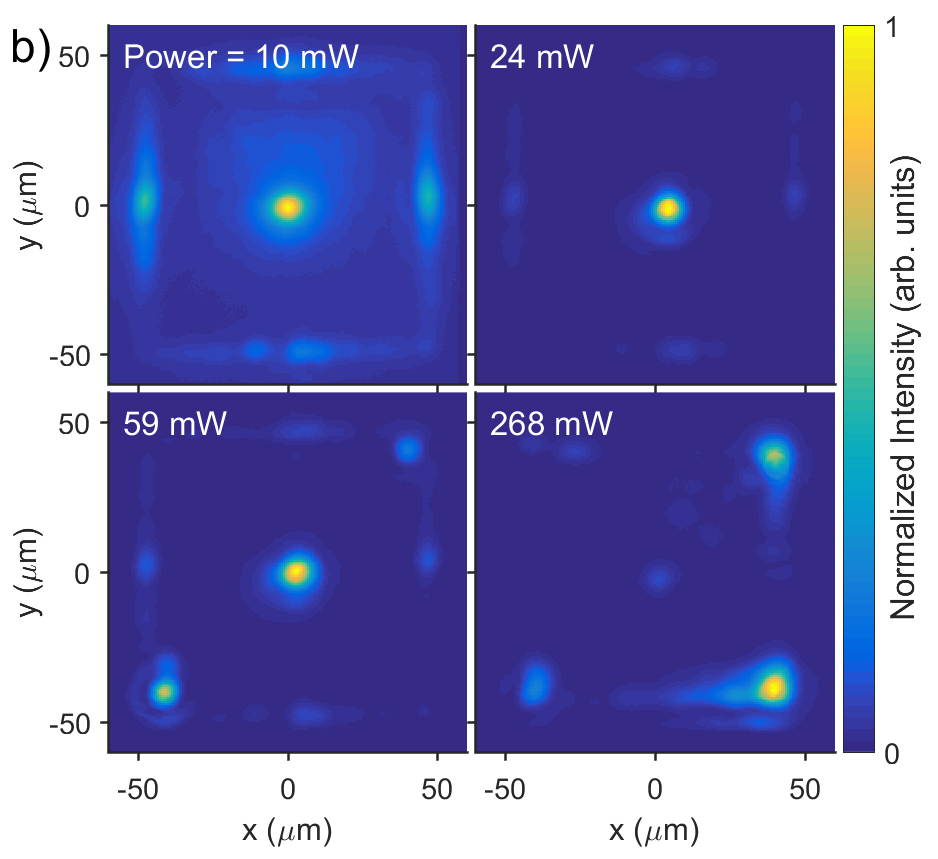}
\end{center}
\caption{Experimental polariton condensate population in the corner traps of the pillar, as a function of incoherent pump power, for Device 3, different from the Device~4 used for the data of Fig.~3 of the main text.}
\label{3thres}
\end{figure}

We review here results also reported in Ref.~\onlinecite{photwest}. Figure~\ref{3thres} shows the condensate intensity in the traps at the corners and edges of the pillar as a function of optical pump power, for one of the devices. There are three thresholds for nonlinear increase of the intensity clearly visible here; in some devices only two of these thresholds are visible. 

This threshold behavior can be understood as follows. Below the lowest threshold, there is a thermal distribution of polaritons concentrated at the pump spot, diffusing throughout the pillar (see Figure~\ref{3thres}(b), 10 ~mW data). The emission at the corners and edges increases linearly with the pump power. The first threshold occurs when condensation occurs at the optical excitation spot in the middle of the pillar. As documented in Refs.~\cite{prx} and \cite{mark-prb} below, in our long-lifetime structures, as the excitation density is increased, a ``quasicondensate'' first appears at the excitation spot, characterized by increased occupancy of low energy states, but without long-range coherence or superfluidity. This emission is spectrally shifted to the blue by the potential energy of the exciton cloud on which it sits. The emission of the quasicondensate is spectrally narrow but still has measurable spectral width, and is far more intense than all other regions (see Figure~\ref{3thres}(b), 24~mW data). Polaritons from this quasicondensate are not trapped (in fact, they may be viewed as anti-trapped, repelled away by the exciton cloud potential); these polaritons stream freely, ballistically, away from the excitation region. Some of these particles accumulate in the low-energy traps due to partial thermalization, but not enough for condensation in these traps. 

A second threshold occurs when there is a ``true''  condensate that appears in the traps at the edges of the pillar, namely the corners and edges of the pillar, which are the lowest energy points \cite{pillar}. This state is characterized by emission which is spectrally very sharp and which is localized to the ground states of the traps (see Figure~\ref{3thres}(b), 59~mW data). This behavior has also been seen in other trapping geometries \cite{prx,gang}. This drop in the photon emission energy, extreme spectral narrowing, and long-distance motion is quite dramatic in the experiments. We tentatively identify this with the onset of superfluid flow into local traps due to collisions of polaritons in streaming out of the quasicondensate at the laser excitation spot.  At high enough density, the streaming polaritons from the excitation region may have a high enough collision rate with each other to allow scattering into lower energy states, and thereby occupy the trap states at lower energy. The condensates in these traps are well behaved, with well-defined energy.

The third threshold has been seen only in these pillars, and occurs when the various trapped condensates in the corners of the pillar begin to spread out of their traps into the regions between the corner traps (see Figure~\ref{3thres}(b), 268~mW data). Eventually, a single monoenergetic condensate, i.e., phase locking of multiple condensates, appears, as reported previously \cite{pillar}. Surprisingly, this the third threshold also leads to a nonlinear increase of the total emission intensity, which appears to be associated with the phase locking of the the separated condensates into a single, macroscopic, or ``global'' condensate.

We examined nine devices, 
with detunings ranging from $ \delta = -10$~meV to around $\delta = -5$~meV.  In several of the devices, the lower two thresholds are so close together that they appear as a single threshold, leading to only two clear thresholds, as in Figure~3b of the main text. The onset of current increase seen in Figure 2 of the main text is associated with the second threshold, when the condensate first appears in the corner traps. 

\section{Electrical Characteristics of the Polariton Devices} 

\begin{figure}
\begin{center}
\includegraphics[width=0.55\textwidth]{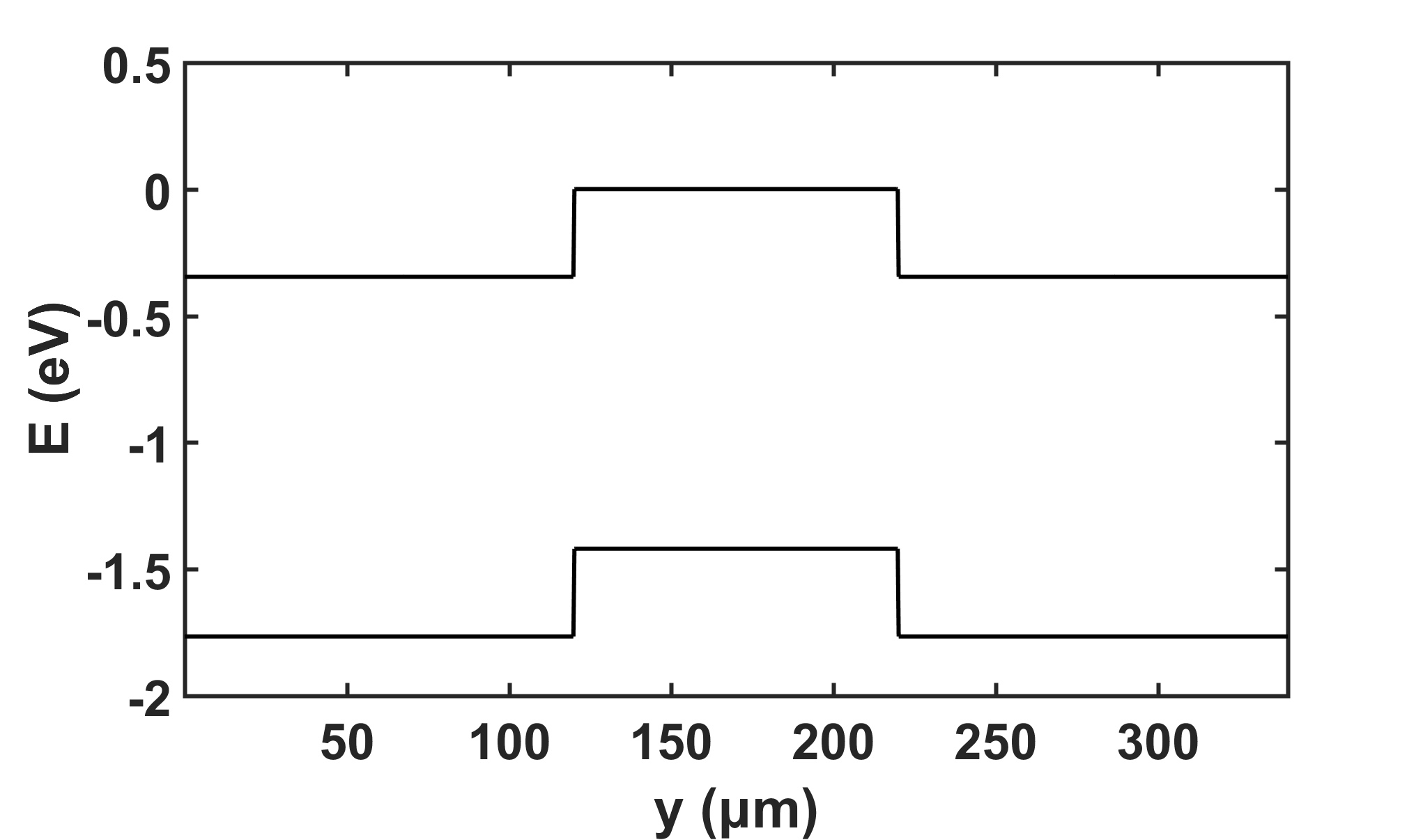}
\end{center}
\caption{Plot of the bands of the  structure, produced by using the ``1D Poisson Beta 8j'' program prepared by Dr.~Gregory Snider at Notre Dame University \cite{snider}, which solves one-dimensional Poisson equations for semiconductor structures.  The heterostructure used here consists of one layer of lightly n-doped GaAs with a doping concentration of $4.3\times10^{13}$ ~cm$^{-3}$ sandwiched between two layers of $n$-doped GaAs with a doping concentration of $1.6\times 10^{19}$~cm$^{-3}$, based on the area densities from the Hall measurements and the estimated thickness. The boundary conditions are considered as ohmic, with no applied bias; the assumes that all shallow dopants are fully ionized. For the lightly doped region, the thickness was 100~$\mu$m, and for the doped contacts, the thickness was 120~$\mu$m.}
\label{bands}
\end{figure}

As discussed in the main text, the devices are nominally $n$-$i$-$n$, with the outer contacts heavily $n$-doped and the region in the middle nominally undoped. We performed Hall measurements for both the contact regions and the nominally undoped optical region; the contact area is $n$-type with area density $6.7\times 10^{15}$~cm$^{-2}$, and the nominally updoped region that holds the polariton condensate is also $n$-type, with area density $1.8\times 10^{10}$ ~cm$^{-2}$. The difference in the doping densities gives a band offset of 0.34 eV, so that each of the contact interfaces acts like a photodiode. The band bending is shown in Figure~\ref{bands}, with a depletion length of 0.1~$\mu$m for the junction between these two doping densities. The structure is effectively a phototransistor, but with a much longer distance (50-100 $\mu$m) between the collector and emitter. Because of this, the photocurrent depends on the voltage drop between the contacts; for low voltage drop, the photogenerated carriers will recombine with each other and not make it to the contacts. 

The Ebers-Moll model for the phototransistor is given by the equations
\begin{eqnarray}
I_B(G) &=& \frac{I_s}{\beta}\left(e^{e(V_B-V_E)/k_BT}-1\right) +  \frac{I_s}{\beta}\left(e^{e(V_B-V_V)/k_BT}-1\right)  \nonumber\\
I_C &=& I_s\left(e^{e(V_B-V_E)/k_BT}-1\right)\left(1-\frac{1}{\beta}\right)-  I_s\left(e^{e(V_B-V_C)/k_BT}-1\right)  
\end{eqnarray}
where $V_E$ and $V_C$ are the emitter and collector voltages, respectively, $\beta$ is the gain factor, and $I_s$ is the saturation current. The photogenerated carriers give a base current that depends on the voltage drop which pulls the electrons and holes away from each other and toward the contacts. We model this simply as $I_B = G(1-e^{-V_{CE}^2/\sigma^2})$. Accounting for series resistance $R$, the measured current will be given by the solution of
\begin{eqnarray}
I_{\rm meas} &=& \frac{V-V_C}{R} = I_C(V_C).
\end{eqnarray}
where $V$ is the applied voltage. 

\begin{figure}
\includegraphics[width=0.65\textwidth]{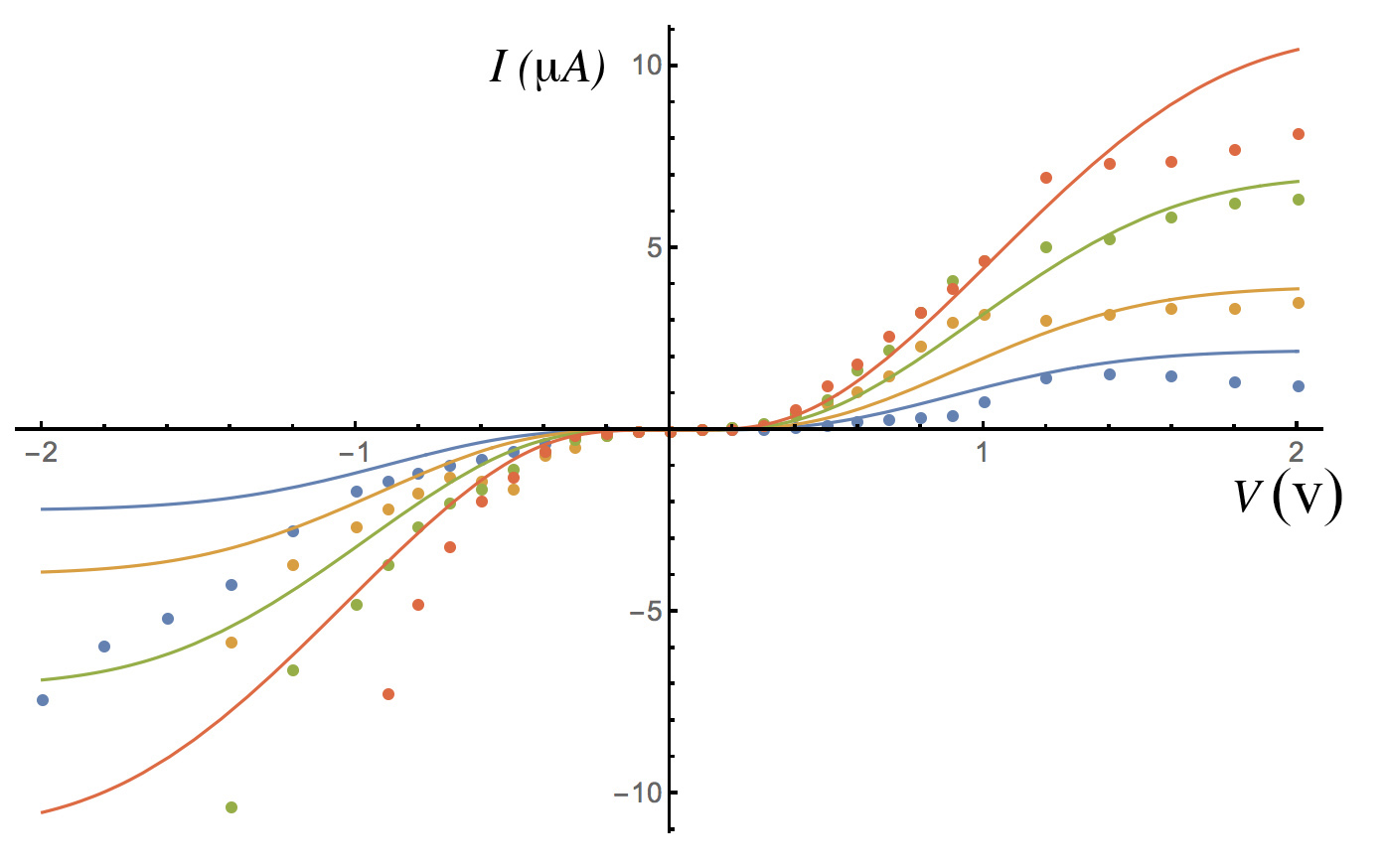}
\caption{Dots: Measured current vs.~voltage for Device 4 at various fixed optical pump powers: 5 mW (blue), 9 mW (yellow), 16 mW (green), and 25 mW (red). Lines: solutions of the Ebers-Moll model discussed in the text, for four illumination intensities with the same ratios as the four powers used for the experimental data.}
\label{IV}
\end{figure}

Figure~\ref{IV} shows the current-voltage data for Device 4 (used in the data of Figure~2 in the main text) for several illumination powers at the excitation spot, compared to solutions of the Ebers-Moll model for the phototransistor given above. As seen in this figure, although the model is symmetric with bias direction, the measured current is not, presumably because the photocurrent is somewhat asymmetric, larger in the negative direction, for this particular case. This could occur, for example, if the distance traveled by the photogenerated holes to a contact is further in one direction than the other, due to the placement of the laser pump spot. 
When there is no illumination of the device, the device has no measurable current flow until very high voltage. 

\begin{figure}
\includegraphics[width=0.6\textwidth]{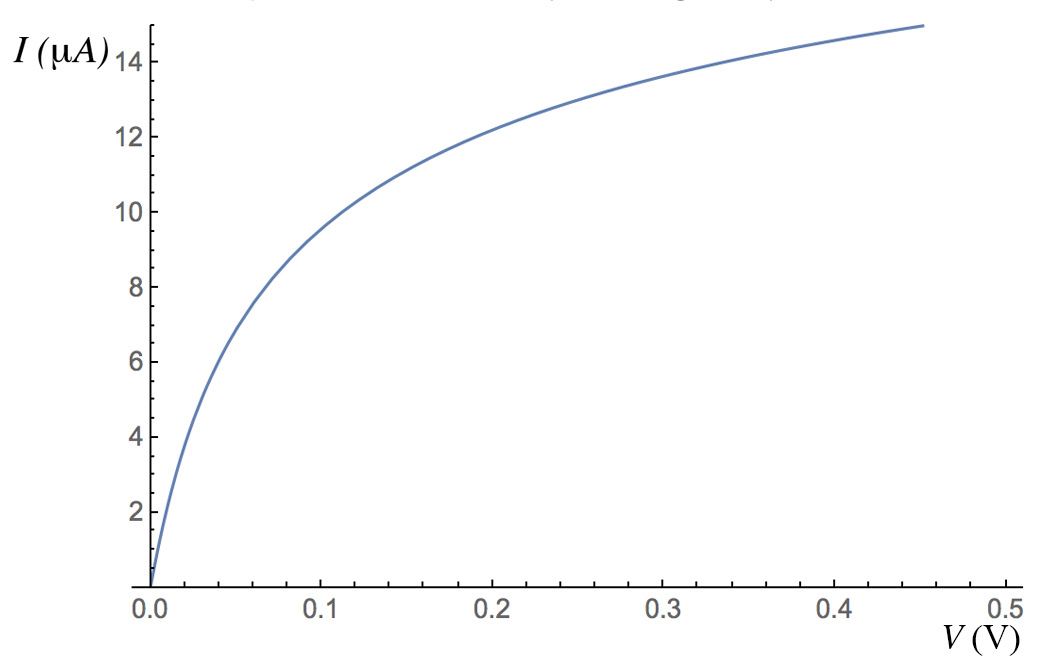}
\caption{Current vs.~pump power for the same Ebers-Moll model and parameters for Device 4 used for Figure~\protect\ref{IV}.}
\label{IP}
\end{figure}

Figure~\ref{IP} shows the behavior seen at low power in Figure 2a of the main text, namely linear growth of the current with illumination power at low illumination, then sublinear behavior and eventually saturation, which occurs when the effective resistance of the device is so small that the current is limited by the current through the middle resistance, $I = V_i/R$.  The Ebers-Moll model gives no way for the current in the phototransistor ever to be superlinear with illumination intensity. As discussed in the main text, superlinear increase only occurs when there is a polariton condensate. 

\appendix

\end{document}